\documentclass[12pt]{article}
\usepackage{amsmath,amsthm,amsfonts}

\input epsf.tex
\newdimen\epsfxsize
\newdimen\epsfysize
\def\qed{\vrule height5pt width3pt depth.5pt}

\theoremstyle{plain}
\newtheorem{thm}{Theorem}[section]

\newtheorem{lem}[thm]{Lemma}

\newtheorem{rem}{Remark}[section]

\begin{document}

\title{Anyonic Topological Quantum Computation and the Virtual Braid Group}

% information for author

\author{H. A. Dye \\
McKendree University\\
hadye@mckendree.edu \\
Louis H. Kauffman \\
University of Illinois at Chicago \\
kauffman@uic.edu}

\maketitle

\begin{abstract} We introduce a recoupling theory for virtual braided trees. This recoupling theory can be utilized to incorporate swap gates into anyonic models of quantum computation.
\end{abstract}
\section{Classical Trees}
Louis Kauffman and Sam Lomonaco reconstructed the Fibonacci model of quantum computation in \cite{lousam} by applying the 2-strand Temperly-Lieb recoupling and the skein relation to braided trees. (The original Fibonacci model is discussed in \cite{presskill}, \cite{freedman}, and \cite{bonesteel}.) In the context of quantum computation, virtual crossings can be regarded as swap gates.

We incorporate virtual crossings into this model as generalized swap gates. This allows us to extend and compute quantum topological invariants to the category of virtual links. In the context of quantum computation, the anyonic model can be used to compute the Jones polynomial of knots and links. Incorporating virtual crossings as generalized swap gates leads to a model that can be used as a quantum algorithm for the Jones polynomial of virtual links.

Virtual crossings can be incorporated into the Fibonacci model, but the calculus described in \cite{lousam} is no longer sufficient to left associate an arbitrarily constructed virtual tree.
In this paper, we extend the graphical calculus to virtual braided trees.

\begin{figure}[htb] \epsfysize = 0.75 in
\centerline{\epsffile{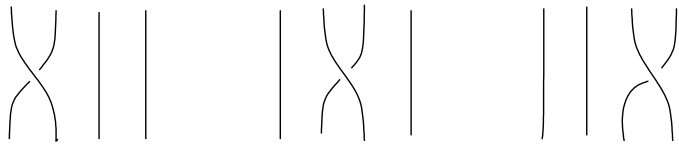}}
\caption{Generators of $B_4 $}
\label{fig:genbraid}
\end{figure}

\begin{figure}[htb] \epsfysize = 0.75 in
\centerline{\epsffile{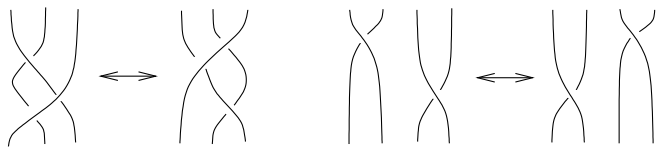}}
\caption{Equivalent braids}
\label{fig:relations}
\end{figure}

Recall Artin's $n$-strand braid group, $B_n$. Let $ \lbrace \sigma_1 , \sigma_2 , \ldots \sigma_{n-1} \rbrace $  denote the generators of $B_n $. The generators of $ B_4 $, the four strand braid group, are illustrated in figure \ref{fig:genbraid}.
The $n$-strand braid group is determined by the following relations on the generators:
\begin{itemize}
\item $ \sigma_i \sigma_{i+1} \sigma_i = \sigma_{i+1} \sigma_i \sigma_{i+1} $
\item $ \sigma_i \sigma_j = \sigma_j \sigma_i $ for $ |i-j|>1 $.
\end{itemize} 
These relations determine equivalent braids, via the Reidemeister II and III moves, as shown in figure \ref{fig:relations}.

\begin{figure}[htb] \epsfysize = 0.75 in
\centerline{\epsffile{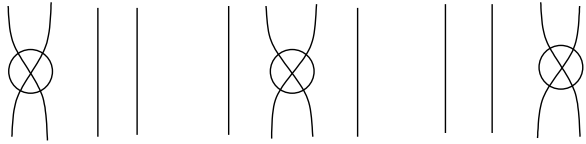}}
\caption{Virtual generators}
\label{fig:virtual}
\end{figure}

\begin{figure}[htb] \epsfysize = 1.5 in
\centerline{\epsffile{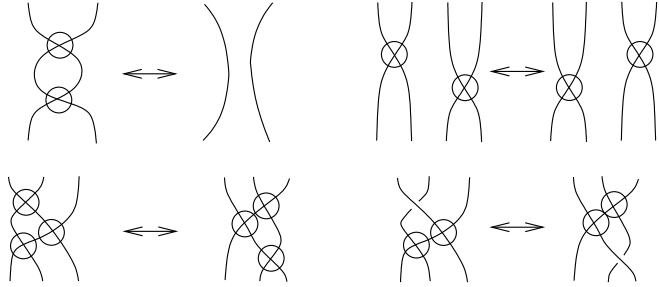}}
\caption{Virtual braid equivalence}
\label{fig:visquare}
\end{figure}

Extending $B_n$ by the symmetric group results in the $n$-strand virtual braid group, $VB_n$ \cite{kamada-braid} \cite{lou-braid}. We incorporate virtual crossings by adding the generators 
$ v_1 , v_2 , \ldots v_{n-1}  $, where $v_i $ is a $n$-strand braid with a single virtual crossing between strand $i-1$ and strand $i$ as shown in figure \ref{fig:virtual}.
Equivalence classes of virtual braids are determined by the following relations.
\begin{itemize}
\item $v_i ^2 = Id $
\item $ v_i v_j = v_j v_i \text{ for } |i-j|> 1$
\item $ v_i v_{i+1} v_i = v_{i+1} v_i v_{i+1} $
\item $ v_{i+1} v_i \sigma_{i+1} = \sigma_i v_{i+1} v_i $.
\end{itemize}
Diagrammatically, these relations are illustrated in figure \ref{fig:visquare}.

Unitary solutions to the Yang-Baxter equation \cite{ybe} determine unitary representations of the braid group. (Recall that these solutions are universal.) We assume that these representations act on a tensor product; each strand of a braid represents a finite dimensional (usually two dimensional) vector space $V$. Then in the context of the braid group, order two gates that switch strands lead to the usual definition of the swap gate. That is, in an $n$-dimensional vector space $V$ with basis, $ \lbrace |v_1>, |v_2> \ldots |v_n> \rbrace $, an element of $ V \otimes V $ is a qudit and a swap gate sends  $ |v_i> \otimes |v_j> $ to $ |v_j> \otimes |v_i> $. This extends the (tensor) representation of the braid group to the virtual braid group. 

However, we are not working within the context of a tensor representation but instead with a representation that acts on vector spaces associated with trees. The version of the swap gate that we study here is an  order two unitary operator. These generalized swap gates also satisfy the relations in the virtual braid group. These gates also have a more complex behavior than the usual definition of a swap gate, a behavior that we will examine later in this paper.

\begin{rem}
A tensor representation of the braid group corresponding to $ V \otimes V \otimes \ldots \otimes V $, where
$V$ is a two dimensional vector space, can be extended with the matrix defined on $ V \otimes V$:
\begin{equation*}
\begin{bmatrix} 1 & 0 & 0 & 0 \\ 0 & 0 & 1 & 0 \\ 0 & 1 & 0 & 0 \\ 0 & 0 & 0 & 1 \end{bmatrix} .
\end{equation*}
This is the usual swap gate.
\end{rem} 

\begin{figure}[htb] \epsfysize = 1 in
\centerline{\epsffile{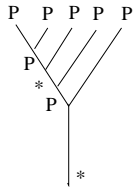}}
\caption{Left associated tree}
\label{fig:lefttree}
\end{figure}

The Fibonacci model is based on the evaluation of associated trees. In a left associated tree, trivalent vertices represent particle interactions and each edge is marked with either a $ P $ or $*$; representing the two possible states of a particle. The uppermost edges are labeled with $P$'s and lower edges are labeled with either a $P$ or a $* $, based on the outcome of a particle interaction. The three possible particle interactions are:
\begin{gather*}
P P = P \\
P P = * \\
* P = P \\
\end{gather*} 
A left associated tree labeled in this fashion is shown in figure \ref{fig:lefttree}.

Since the upper leaves of the tree are labeled with $P$, it follows that $**$ is an impossible sequence of labels.

The space of the left associated trees with $n+2$ upper branches has dimension $f_{n} $ (where $f_n$ represents the $n^{th} $ Fibonacci number). Using the graphical calculus from the Temperly-Lieb algebra \cite{t-lalg}, \cite{lousam}, any classical, braided tree can be transformed into a left associated tree. However, the operations provided by this graphical calculus are not sufficient to transform a virtually braided tree into a left associated tree. This is the problem that concerns us in this paper.

\begin{figure}[htb] \epsfysize = 2 in
\centerline{\epsffile{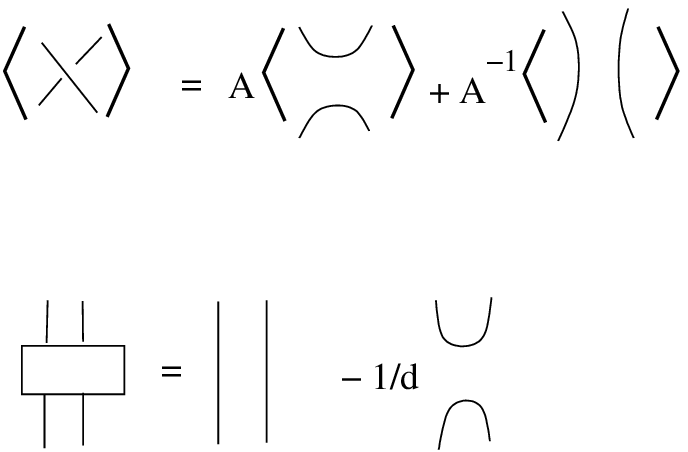}}
\caption{Skein relation and 2-strand recoupler}
\label{fig:graphcalc1}
\end{figure}

\begin{figure}[htb] \epsfysize = 1.5 in
\centerline{\epsffile{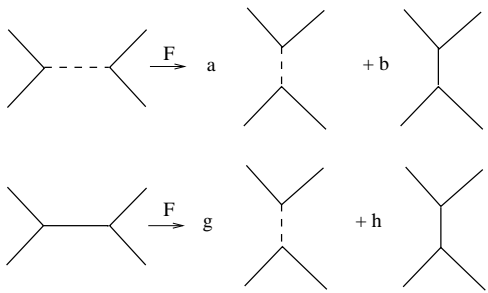}}
\label{fig:ftrans}
\caption{The F transformation}
\end{figure}

The skein relation and the 2-strand recoupler from the Temperly-Lieb algebra are illustrated in figure \ref{fig:graphcalc1}.
These two equations form the basis for all calculations in the remainder of this paper. Here $ d= -A^2 -A^{-2} $.
\begin{figure}[htb] \epsfysize = 2 in
\centerline{\epsffile{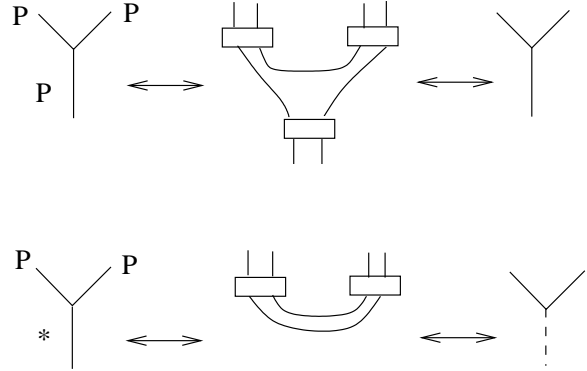}}
\caption{Trivalent vertices and corresponding recoupling diagrams}
\label{fig:trivalent}
\end{figure}
Labeled trees correspond to elements of the $2$-strand Temperly-Lieb recoupling theory as shown in figure \ref{fig:trivalent}.
In this context, an edge labeled with $P$ corresponds to an edge consisting of two strands and an edge labeled with $*$ corresponds to edge with zero strands (that is, the edge does not exist except as a placeholder). Graphically, the edges labeled with $*$ may be represented with a dashed edge. The left association of any classical braided tree only requires two operations: the F transformation and the R transformation. The F transformation is a linear transformation from the set of labeled trees to the set of labeled trees. 
Recall \cite{lousam} and let 
\begin{gather*}
 \Delta = d^2 -1 \\  \Theta  = \frac{(d^2 -1) (d^2 -2)}{d} \\   T= \frac{2(d^2-1)^2 (d^2 -2) \Theta}{d^3}.
\end{gather*}
Here $ \Delta $ is the evaluation of a loop with a projector, $ \Theta $ is the evaluation of a $ \theta $-graph involving trivalent vertices, and $T$ is the evaluation of a tetrahedral graph with trivalent vertices.  
An F transformation results in a linear combination of labeled trees, as shown in figure \ref{fig:ftrans}.
Here $ a = \frac{1}{\Delta} $, $b= \frac{ 1}{ \sqrt{\Delta}} $, $g= \frac{1}{ \sqrt{\Delta}} $, 
and $ h= \frac{-1}{\Delta}$, see \cite{lousam}.
From the point of view of linear algebra, each labeled tree is a basis vector in a complex vector space associated with a single, unlabeled tree. In the case of those trees which standardly have $P$'s labeling their upper branches, the basis vectors consist of all legal labelings of the other edges. An unlabeled tree represents the vector space of all possible labelings of the tree. The F transformation is a unitary linear transformation from the vector space of one tree to the other tree obtained by performing the replacement indicated in figure \ref{fig:ftrans}.

\begin{figure}[htb] \epsfysize = 1 in
\centerline{\epsffile{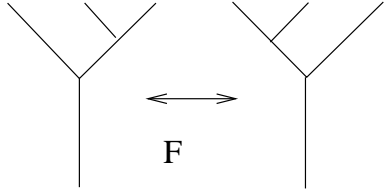}}
\caption{F Transformation}
\label{fig:recoupling}
\end{figure}

Suppressing the information about the labels and coefficients, we can describe the F transformation graphically as shown in figure \ref{fig:recoupling}.

\begin{figure}[htb] \epsfysize = 1.5 in
\centerline{\epsffile{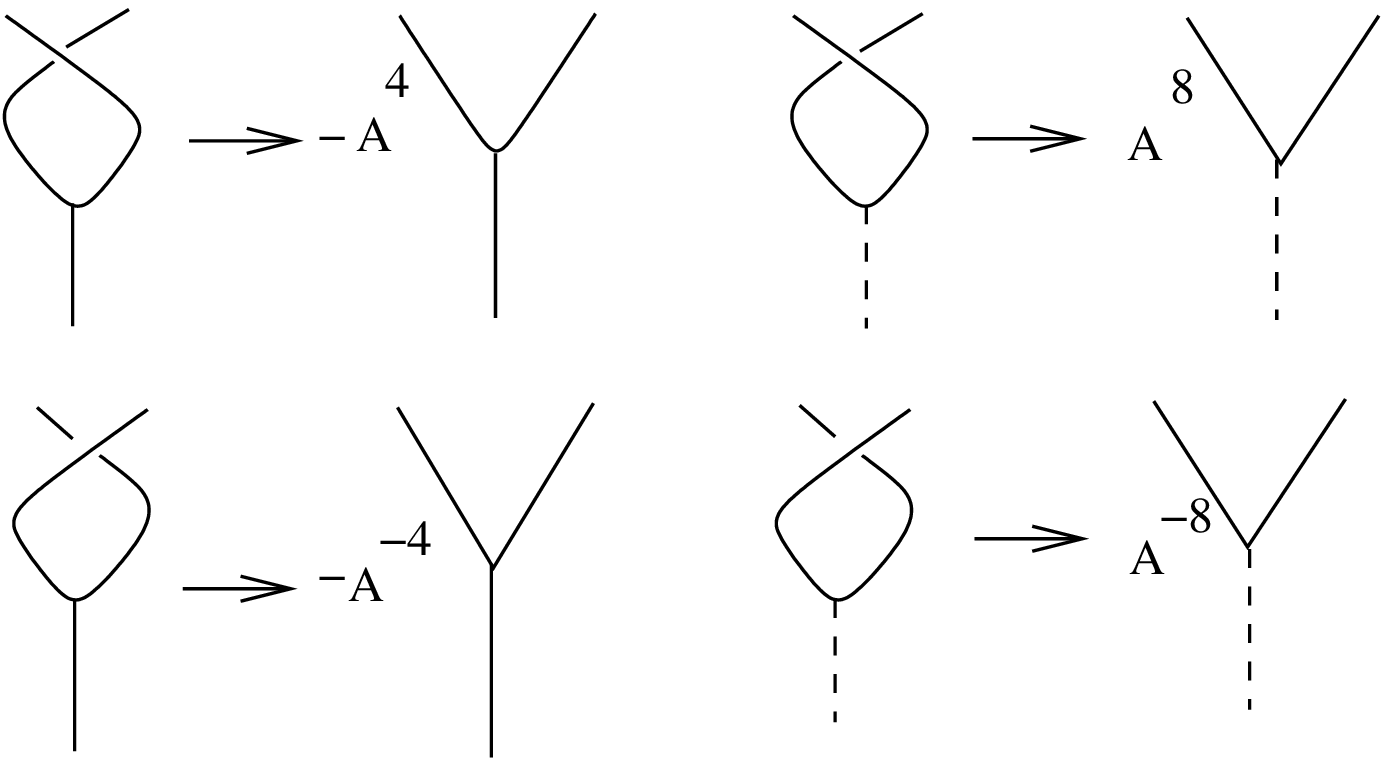}}
\caption{The R transformation}
\label{fig:requiv}
\end{figure}

Similarly, we can apply the R transformation to two adjacent branches with a classical crossing and obtain a multiple of a tree without twisting as shown in figure \ref{fig:requiv}. 

\begin{rem}
As in \cite{lousam}, we describe $ F $ and $R$ by the matrices $F$ and $R$.
\begin{gather}
F=  \begin{bmatrix} \frac{1}{\Delta} & \frac{ 1 }{\sqrt{\Delta} } \\
 \frac{ 1}{ \sqrt{\Delta }} & \frac{-1}{ \Delta } \end{bmatrix} \\
R= \begin{bmatrix} A^{8} & 0 \\
 0 & -A^{4} \end{bmatrix} 
\end{gather}

\end{rem}

\section{Virtual Braided Trees}
\begin{figure}[htb] \epsfysize = 1 in
\centerline{\epsffile{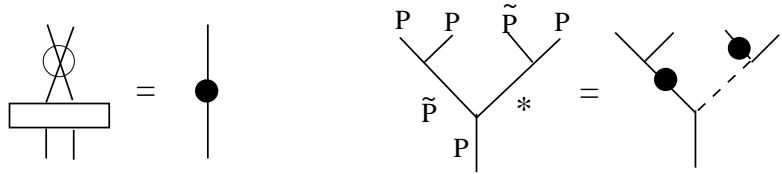}}
\caption{Correspondence between labels and marked edges}
\label{fig:zz}
\end{figure}

In models of quantum computation, swap gates exchange two particles. The effect of a swap gate on two particles is: 
\begin{gather*}
P \otimes P \rightarrow P \otimes P \\
 P \otimes * \rightarrow * \otimes P \\
 * \otimes P \rightarrow P \otimes * \\
 * \otimes * \rightarrow * \otimes * .
\end{gather*}
In virtual braided trees, a line with a dot corresponds to two edges with a virtual crossing leading into a projector. A line of this type is labeled with a $ \tilde{P} $ as shown in figure \ref{fig:zz}.

\begin{figure}[htb] \epsfysize = 1 in
\centerline{\epsffile{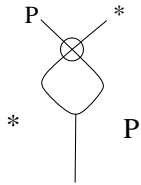}}
\caption{Virtual crossing acting as a swap gate}
\label{fig:virtualvertex}
\end{figure}

\begin{figure}[htb]\epsfysize = 2 in
\centerline{\epsffile{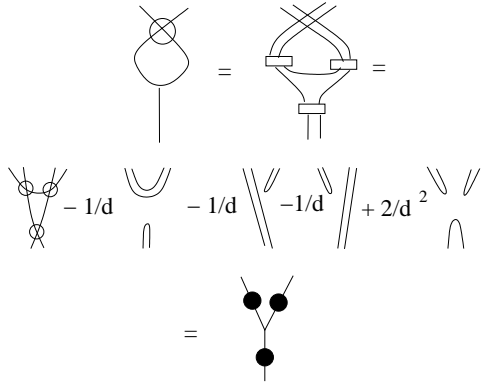}}
\caption{Expanding a trivalent vertex}
\label{fig:fullrewrite}
\end{figure}

A swap gate (realized as a virtual crossing) is illustrated graphically in figure \ref{fig:virtualvertex}. 
We expand a trivalent vertex with a virtual crossing in figure \ref{fig:fullrewrite}.

Resolving the virtual crossing results in 2 strand edges with a virtual crossing between the edges. This represents a third particle type, referred to as $ \tilde{P} $ and indicated in the graphs with a solid dot. We realize that virtual crossings effect not only the coefficient, but also the labeling of the tree. The particles interact as follows:
\begin{gather*}
 PP \rightarrow P, *, \text{ or } \tilde{P} \\
 P \tilde{P} \rightarrow P, *, \text{ or } \tilde{P} \\ 
 P * \rightarrow P  \text{ or } \tilde{P} \\
 \tilde{P} * \rightarrow \tilde{P} \text{ or } P. 
\end{gather*}
This illustrates the more complex behavior of our generalized swap gates.
In the virtual case, left association may alter the labels on the edges of the tree. That is, given an arbitrarily constructed virtual tree with all branches labeled $P$ then, after left association, it is possible to obtain a tree where the upper branches are labeled with $P$ or $ \tilde{P}$. This does not occur in the classical case.
In the following, we develop a calculus that may be applied in order to left associate a virtual tree, resulting in the following theorem.

\begin{figure}[htb] \epsfysize = 1 in
\centerline{\epsffile{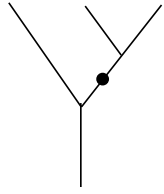}}
\caption{Tree with $ \tilde{P} $ labeled edge }
\label{fig:base}
\end{figure}

\begin{figure}[htb] \epsfysize = 1.5 in
\centerline{\epsffile{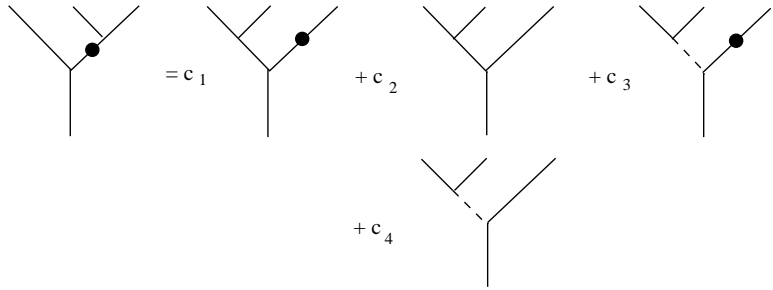}}
\caption{Sum of left associated trees}
\label{fig:leftassoc}
\end{figure}

\begin{thm} \label{the} Any virtual, braided tree can be reformulated as a left associated tree with labels $P$, $*$, and $ \tilde{P}$. \end{thm}
In order to prove this theorem, we will first prove a sequence of lemmas.
First, we determine how to express the tree with an edge labeled $ \tilde{P}$ shown in \ref{fig:base} as a linear combination of labeled, left associated trees.
\begin{lem} We can left associate  a tree with an edge labeled $ \tilde{P}$. The virtual tree shown on the left hand side of figure \ref{fig:leftassoc} is equivalent to the sum of left associated trees shown on the right hand side. \end{lem} 

\begin{figure}[htb]
\epsfysize = 1 in
\centerline{\epsffile{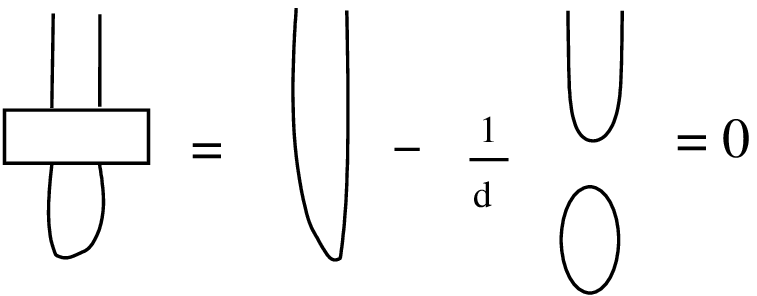}}
\caption{The turnback calculation}
\label{fig:turn}
\end{figure}

\begin{figure}[htb] 
\epsfysize = 1 in
\centerline{\epsffile{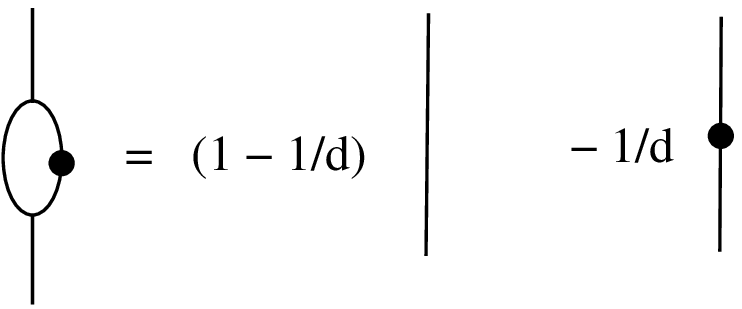}}
\caption{The bubble calculation}
\label{fig:bub}
\end{figure}

\textbf{Proof:} We will use the turnback calculation, figure \ref{fig:turn}, and the bubble calculation, figure
\ref{fig:bub}.

\begin{figure}[htb] \epsfysize = 3.5 in
\centerline{\epsffile{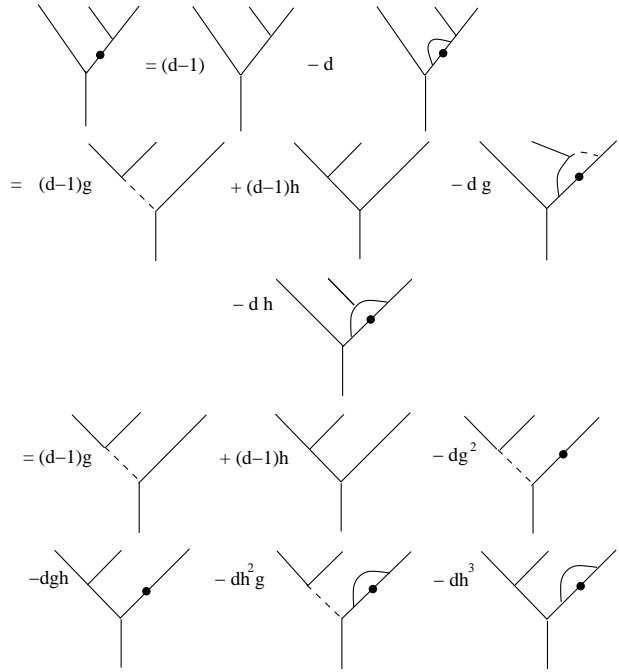}}
\caption{Left association of a virtual tree}
\label{fig:basecalc}
\end{figure}

Applying the bubble calculation to the tree shown in figure \ref{fig:base}, we determine the first equality shown in 
figure \ref{fig:basecalc}.
We then apply the F transformation until we obtain a linear combination of trees that are left associated, with the exception of the bubble on the left hand branch. We then rewrite the linear combination of trees using the bubble calculation to obtain the linear combination shown in figure \ref{fig:leftassoc}.
The coefficients of the trees shown in figure \ref{fig:leftassoc} are:
\begin{align*}
c_1 &= h^3  -dgh & \qquad c_2 &= (d-1)(h-h^3) \\
c_3 &= h^2 g -d g^2 & \qquad c_4 &= (d-1)(g-gh^2). \\
\end{align*}\qed

\begin{figure}[htb] \epsfysize = 1 in
\centerline{\epsffile{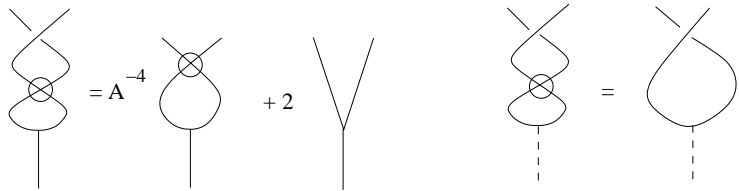}}
\caption{Equivalent sums of trees}
\label{fig:doubletwist}
\end{figure}

\begin{lem}We can left associate a tree that contains a trivalent vertex with virtual and classical crossings as shown in 
figure \ref{fig:doubletwist}.  \end{lem}

\begin{figure}[htb] \epsfysize = 1.5 in
\centerline{\epsffile{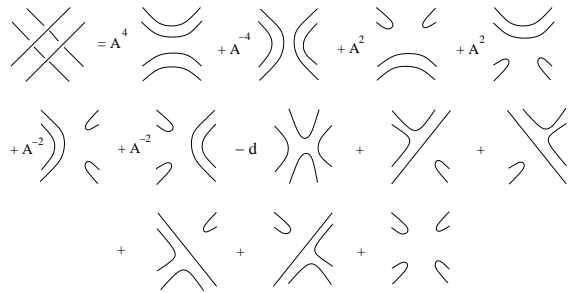}}
\caption{Expanding a double crossing with the skein relation}
\label{fig:doubleexpand}
\end{figure}

\begin{figure}[htb] \epsfysize = 1 in
\centerline{\epsffile{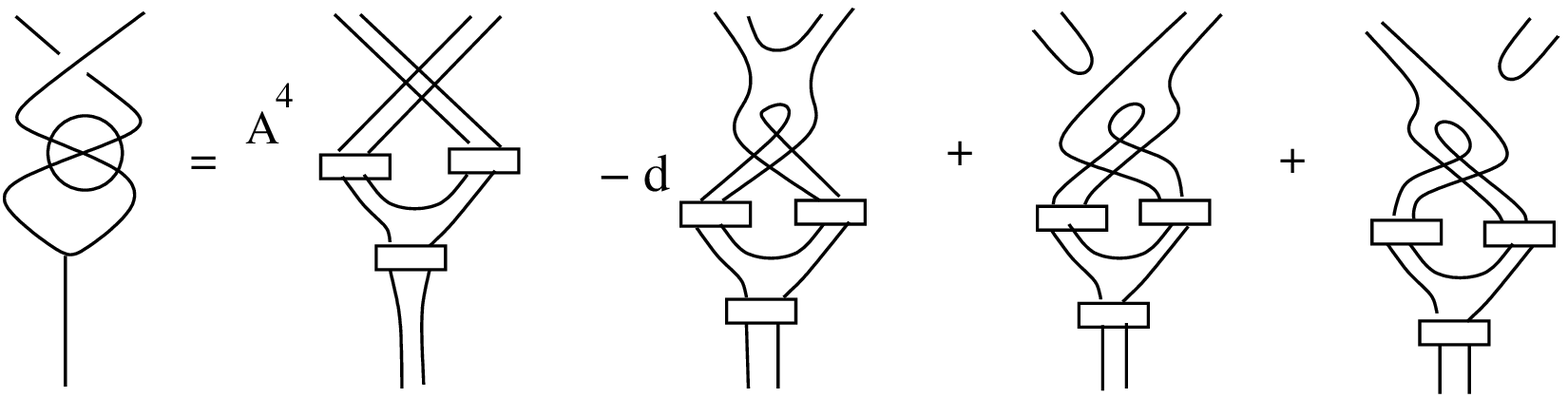}}
\caption{Applying the turnback calculation to the expanded crossing}
\label{fig:dt1}
\end{figure}

\textbf{Proof:}  The tree on the right hand side of figure \ref{fig:doubletwist} is equivalent to a classical tree to which we can apply the R transformation. The tree shown on the left hand side of figure \ref{fig:doubletwist} can be expressed as a linear combination of two trees after some calculations.
We obtain this result by fully expanding the classical crossing with the skein relation shown in figure \ref{fig:doubleexpand}.
We then use the turnback calculation and realize that many of the terms reduce to zero, resulting in the equation shown in 
figure \ref{fig:dt1}.\qed

\begin{figure}[htb] \epsfysize = 1 in
\centerline{\epsffile{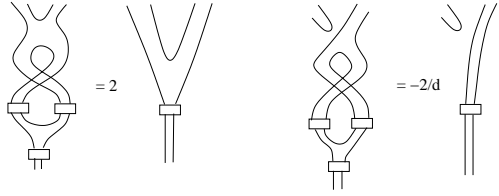}}
\caption{Identities obtained by expanding recouplers}
\label{fig:inter}
\end{figure}
\begin{figure}[htb] \epsfysize = 1 in
\centerline{\epsffile{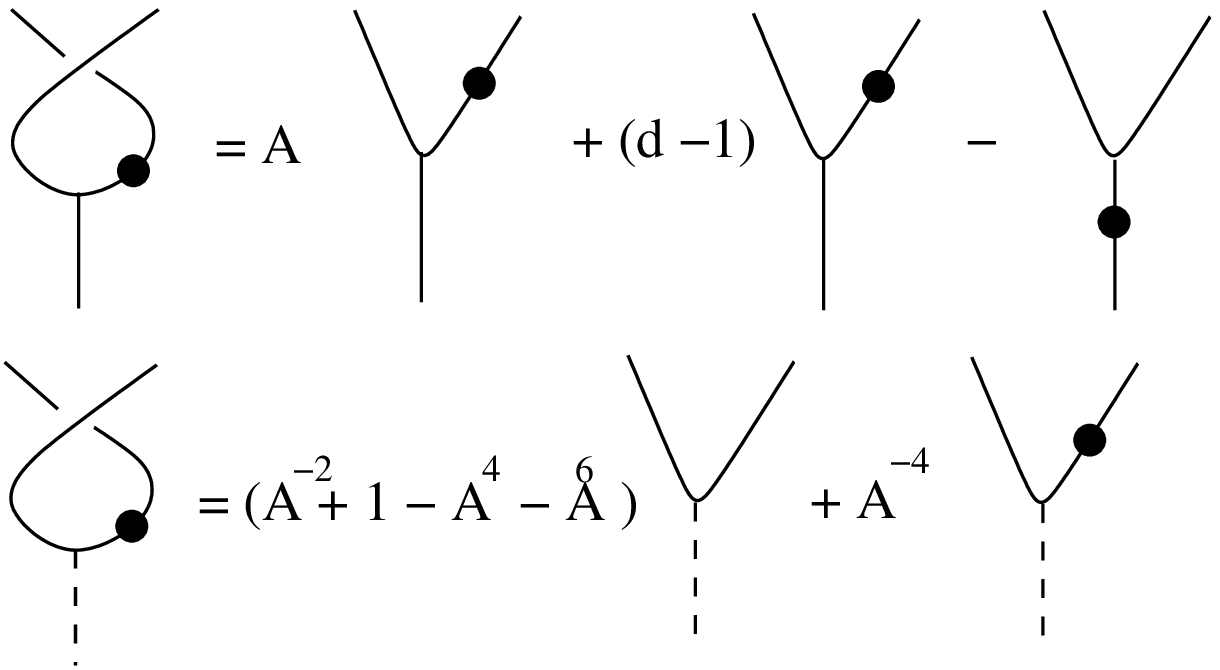}}
\caption{Trees with label $ \tilde{P} $ and a classical crossing}
\label{fig:lastcase}
\end{figure}

\begin{lem}The trees shown on the left hand side of figure  \ref{fig:lastcase} (with classical crossing and branch labeled $ \tilde{P} $) can be left associated.\end{lem}
\textbf{Proof:}  Expanding the recouplers in figure \ref{fig:dt1}, we obtain the equivalences shown in figure \ref{fig:inter}.
From this computation, we obtain the equivalence shown in figure \ref{fig:doubletwist}.
We need only determine how to left associate trees with a classical crossing and an edge labeled $ \tilde{P}$ as shown in figure \ref{fig:lastcase}. These relations are determined by using the expansion shown in figure \ref{fig:doubleexpand} and simplifying. \qed

These additional operations, combined with the F transformation and R transformation, are sufficient to left associate all virtual, braided trees. We have proved theorem \ref{the}.
In order to utilize this graphical calculus in the context of models of quantum computation, we must extend the basis and  determine matrix representations of these relations and the dimension of the space of left associated virtual trees. These linear transformations relating trees must
also be described as unitary matrices. We will explore these linear representations in a subsequent paper. 
Note that the calculus that we have constructed so far contains new order two gates that fit into a representation of the virtual braid group. These gates (being of order two) can be regarded as generalizations of a simple swap gate. The use we intend for these swap gates are for extensions of topological quantum invariants and the Jones polynomial to virtual links. The gates may be of use in topological quantum computing in general since the original Fibonacci model is universal for quantum computing.


\begin{thebibliography}{10}

\baselineskip=12pt % space between lines
\parskip=2pt plus 1pt % space between paragraphs

\bibitem{bonesteel}
\newblock{ Bonesteel, N. E.; Hormozi, L.; Zikos, G.; Simon, S. H. Braid topologies for quantum computation. Phys. Rev. Lett. 95 (2005), no. 14, 140503, 4 pp. 81P68}


\bibitem{brylinski} 
\newblock{ Brylinski, Jean-Luc; Brylinski, Ranee Universal quantum gates. Mathematics of quantum computation, 101--116, Comput. Math. Ser., Chapman \& Hall/CRC, Boca Raton, FL, 2002.}

\bibitem{ybe}
\newblock{H. A. Dye, Unitary solutions to the Yang-Baxter equation in dimension four, 
Quantum Information Processing,  2 (2003), no. 1-2, 117--150}


\bibitem{wrt}
\newblock{H. A. Dye and Louis H. Kauffman, Virtual knot diagrams and 
the Witten-Reshetikhin-Turaev invariant,}
\newblock{The Journal of Knot Theory and its Ramifications, 14 (2005), no. 8, 1045--1075}


\bibitem{freedman}
\newblock{Freedman, Michael H.; Kitaev, Alexei; Wang, Zhenghan Simulation of topological field theories by quantum computers. Comm. Math. Phys. 227 (2002), no. 3, 587--603.}

\bibitem{kamada-braid}
\newblock{Kamada, Seiichi Invariants of virtual braids and a remark on left stabilizations and virtual exchange moves. Kobe J. Math. 21 (2004), no. 1-2, 33--49.}

\bibitem{lousam}
\newblock{Kauffman, Louis H.; Lomonaco, Samuel J., Jr. $q$-deformed spin networks, knot polynomials and anyonic topological quantum computation. J. Knot Theory Ramifications 16 (2007), no. 3, 267--332.}


\bibitem{t-lalg}
\newblock{Kauffman, Louis H.; Lins, S$\acute{o}$stenes L. Temperley-Lieb recoupling theory and invariants of $3$-manifolds. Annals of Mathematics Studies, 134. Princeton University Press, Princeton, NJ, 1994.}

\bibitem{lou-braid}
\newblock{Kauffman, Louis H.; Lambropoulou, Sofia. Virtual braids. Fund. Math. 184 (2004), 159--186. }

\bibitem{presskill}
\newblock{Preskill, J. Topological computing for beginners, (slide presentation), Lecture Notes for Chapter 9 -- Physics 219 -- Quantum Computation. \textit{ http://www.iqi.caltech.edu/$\equiv$preskill/ph219}.}



\end{thebibliography}
\end{document}